\documentclass{article}
\usepackage{slashed,verbatim,subfigure}
\usepackage[numbers,sort&compress]{natbib}
\usepackage{amsmath}
\usepackage{amssymb}
\usepackage{appendix}
\usepackage{graphicx}
\usepackage{dcolumn}
\usepackage{bm}
\usepackage[colorlinks=true,linktocpage=true,
linkcolor=blue,citecolor=blue]{hyperref}
\usepackage[a4paper]{geometry}
\newcommand{\be}{\begin{eqnarray}}
\newcommand{\ee}{\end{eqnarray}}
\newcommand{\ec}{\sigma_{\rm el}}

\begin{document}
\large
\title{\bf{Effect of magnetic field on the charge and thermal transport 
properties of hot and dense QCD matter}}
\author{Shubhalaxmi Rath\footnote{srath@ph.iitr.ac.in}~~and~~Binoy Krishna 
Patra\footnote{binoy@ph.iitr.ac.in}\vspace{0.03in} \\ 
Department of Physics, Indian Institute of Technology Roorkee, Roorkee 247667, India}
\date{}
\maketitle
\begin{abstract}
We have studied the effect of strong magnetic field on the 
charge and thermal transport properties of hot QCD matter 
at finite chemical potential. For this purpose, we have 
calculated the electrical conductivity ($\sigma_{\rm el}$) 
and the thermal conductivity ($\kappa$) using kinetic theory 
in the relaxation time approximation, where the 
interactions are subsumed through the 
distribution functions within the quasiparticle model at 
finite temperature, strong magnetic field and finite 
chemical potential. This study helps to understand the 
impacts of strong magnetic field and chemical potential on the 
local equilibrium by the Knudsen number ($\Omega$) through 
$\kappa$ and on the relative behavior between thermal conductivity and 
electrical conductivity through the Lorenz number ($L$) 
in the Wiedemann-Franz law. We have observed that, both 
$\sigma_{\rm el}$ and $\kappa$ get increased in the presence of strong 
magnetic field, and the additional presence of chemical 
potential further increases their magnitudes, where $\sigma_{\rm el}$ 
shows decreasing trend with the temperature, opposite to its 
increasing behavior in the isotropic medium, whereas 
$\kappa$ increases slowly with the temperature, contrary to 
its fast increase in the isotropic medium. The variation 
in $\kappa$ explains the decrease of the Knudsen number 
with the increase of the temperature. However, in the presence 
of strong magnetic field and finite chemical potential, $\Omega$ 
gets enhanced and approaches unity, thus, the system may move slightly away 
from the equilibrium state. The Lorenz number ($\kappa/(\sigma_{\rm el} T))$ 
in the abovementioned regime of strong magnetic field and finite 
chemical potential shows linear enhancement with the temperature 
and has smaller magnitude than the isotropic one, thus, it 
describes the violation of the Wiedemann-Franz law for the hot and 
dense QCD matter in the presence of a strong magnetic field. 

\end{abstract}

\newpage

\section{Introduction}
At high temperatures and/or chemical potentials the 
system can transit to a state consisting of deconfined 
quarks and gluons, called as quark-gluon plasma (QGP) 
and such conditions are evidenced in ultrarelativistic 
heavy-ion collisions at Relativistic Heavy Ion Collider 
(RHIC) \cite{Arsene:NPA757'2005}, Large Hadron Collider 
(LHC) \cite{Carminati:JPG30'2004}, and are expected to 
be produced in the Compressed Baryonic Matter (CBM) 
experiment at Facility for Antiproton and Ion Research 
(FAIR) \cite{Senger:CEJP10'2012}. In addition, the 
collision for nonzero impact parameter also 
produces a strong magnetic field, whose magnitude varies 
from $eB=m_{\pi}^2$ ($\simeq 10^{18}$ Gauss) at RHIC to 15 $m_{\pi}^2$ at 
LHC \cite{Skokov:IJMPA24'2009}. Although the quark chemical 
potential is very small in the initial stages 
of ultrarelativistic heavy ion collisions, it is not zero. 
Some studies also suggest that at temperature around 160 MeV, the 
baryon chemical potential is approximately 300 MeV \cite{P:JPG28'2002,Cleymans:JPG35'2008,Andronic:NPA837'2010}. 
Especially in the strong 
magnetic field regime, the baryon chemical potential is observed to 
go up from 0.1 GeV to 0.6 GeV \cite{Fukushima:PRL117'2016}, 
which also implies an increase in the quark chemical potential. 
In recent years, a lot of observations have been made to 
explore the effect of strong magnetic field on the properties 
of hot QCD matter, such as the thermodynamic and magnetic 
properties \cite{Bandyopadhyay:PRD100'2019,
Rath:JHEP1712'2017,Expansion,Karmakar:PRD99'2019}, the chiral 
magnetic effect \cite{Fukushima:PRD78'2008,Kharzeev:NPA803'2008}, the 
dilepton production from QGP \cite{Tuchin:PRC88'2013,Mamo:JHEP1308'2013} etc. 

It has been observed that, in a model where the hot QCD medium is 
approximated as a gas of hadron resonances, the increase of the 
electric charge fluctuation with increasing magnetic field is 
significant at finite chemical potential \cite{Fukushima:PRL117'2016} 
and according to the field-theoretical calculation in the presence of 
magnetic field \cite{Fukushima:JHEP2004'2020}, the increase in the 
electric charge fluctuation suggests significant enhancement in the 
longitudinal electrical conductivity at finite quark 
chemical potential. Thus, it would be interesting to see 
how different transport coefficients of hot QCD matter behave at 
finite chemical potential in the presence of strong magnetic field. So, 
in this work, we have calculated the electrical conductivity ($\ec$) 
and the thermal conductivity ($\kappa$) in the presence of both strong 
magnetic field and chemical potential and compared them with their 
counterparts in the absence of magnetic field and chemical potential. 
After understanding these transport properties, we have further studied the 
effects of strong magnetic field and chemical potential on the local 
equilibrium by the Knudsen number ($\Omega$) through the thermal 
conductivity and on the relative behavior between electrical conductivity 
and thermal conductivity by the Lorenz number ($L$) in Wiedemann-Franz law. 
In the presence of an external magnetic field, the dispersion relation 
of a $f$th flavor of quark with absolute charge $|q_f|$ and mass $m_f$ 
becomes modified as $\omega_{f,n}=\sqrt{p_L^2+2n|q_fB|+m_f^2}$, where the 
motion along the longitudinal direction ($p_L$) 
(with respect to the direction of magnetic field) resembles with that 
of a free particle and the motion along the transverse 
direction ($p_T$) is expressed in terms of the Landau levels ($n$). 
In the strong magnetic field (SMF) limit 
($|q_fB| \gg T^2$ and $|q_fB| \gg m_f^2$), the charged 
particles can not jump to the higher Landau levels 
due to very high energy gap $\sim\mathcal{O}(\sqrt{|q_fB|})$, 
so only the lowest Landau level (LLL) is populated. Thus, the 
motion of charged particle along the longitudinal direction is 
much greater than the motion along the transverse direction, 
{\em i.e.} $p_L \gg p_T$ and this develops an anisotropy in the momentum space. 

According to some recent observation, the medium is formed at the 
similar time scale of the production of strong magnetic field due to 
faster thermalization. The electrical conductivity of QGP plays an 
important role in extending the lifetime of such strong magnetic field 
\cite{Tuchin:AHEP2013'2013,Conductivities}. Thus, the transport properties of the 
medium are expected to be significantly affected by the strong magnetic 
field. One of the transport coefficients is the electrical conductivity, because 
of which, electric current is produced in the early stage of the heavy-ion 
collision and its value is also important for the strength of chiral 
magnetic effect \cite{Fukushima:PRD78'2008}. Moreover, the strength of the 
charge asymmetric flow in mass asymmetric collisions is given by the 
electrical conductivity \cite{Hirono:PRC90'2014}. Different models have 
previously investigated the influence of magnetic field on 
$\sigma_{\rm el}$, such as the quenched SU($2$) lattice 
gauge theory \cite{Buividovich:PRL105'2010}, the dilute instanton-liquid 
model \cite{Nam:PRD86'2012}, the nonlinear electromagnetic currents \cite{Kharzeev:PPNP75'2014,Satow:PRD90'2014}, the axial Hall current 
\cite{Pu:PRD91'2015}, the real time formalism using the diagrammatic method 
\cite{Hattori:PRD94'2016}, the effective fugacity approach 
\cite{Kurian:PRD96'2017} etc. Another transport coefficient is the thermal 
conductivity, that is associated with the heat flow or thermal 
diffusion in a medium. According to the estimation of effective fugacity 
model \cite{Kurian:EPJC79'2019}, $\kappa$ becomes larger in the presence 
of magnetic field. In this work, we have followed the kinetic theory 
approach to calculate the electrical and thermal conductivities in the 
presence of strong magnetic field and finite chemical potential. 

Using the thermal conductivity, we intend to observe the 
effects of strong magnetic field and chemical 
potential on the local hydrodynamic equilibrium of the 
medium through the Knudsen number ($\Omega$). The Knudsen number 
is defined as the ratio of the mean free path ($\lambda$) to the 
characteristic length ($l$) of the system, where $\lambda$ is defined 
in terms of $\kappa$ as $\lambda=3\kappa/(vC_V)$, with $v$ and $C_V$ 
represent the relative speed and the specific heat at constant volume, 
respectively. For the validity of the hydrodynamic equilibrium, the 
mean free path needs to be smaller than the characteristic length of the 
system. The relative importance of the thermal and electrical 
conductivities can be understood through the Wiedemann-Franz law, which 
states that the ratio of the electronic contribution 
of the thermal conductivity to the electrical conductivity 
($\kappa/\sigma_{\rm el}$) is the product of Lorenz number 
and temperature. This law is well satisfied by the metals, because they 
are good heat and electrical conductors. However, different 
systems also report the violation of the Wiedemann-Franz law, 
such as the thermally populated electron-hole plasma in 
graphene \cite{Crossno:S351'2016}, the two-flavor quark matter 
in the Nambu-Jona-Lasinio model \cite{Harutyunyan:PRD95'2017}, the strongly 
interacting QGP medium \cite{Mitra:PRD96'2017}, the unitary Fermi gas \cite{Husmann:PNAS115'2018,Han:PRA100'2019} and the hot hadronic matter \cite{Rath:EPJA55'2019}. Now it is interesting to see 
how the Lorenz number, $L=\kappa/(\sigma_{\rm el}T)$ 
in the Wiedemann-Franz law becomes influenced by the strong 
magnetic field and finite chemical potential. 

In this work, we have studied the charge and thermal transport 
properties and their applications through the Knudsen number 
and the Lorenz number in the presence of strong magnetic field and 
finite chemical potential and observed how these properties are 
different from their respective behaviors in a medium in the 
absence of magnetic field and chemical potential. The electrical 
conductivity and the thermal conductivity of the hot QCD medium can 
be calculated using different approaches, {\em viz.}, the 
relativistic Boltzmann transport equation \cite{Muronga:PRC76'2007,Puglisi:PRD90'2014,Thakur:PRD95'2017,
Yasui:PRD96'2017}, the Chapman-Enskog approximation \cite{Mitra:PRD94'2016,Mitra:PRD96'2017}, the 
correlator technique using Green-Kubo formula \cite{Nam:PRD86'2012,Greif:PRD90'2014,Feng:PRD96'2017}, the lattice 
simulation \cite{Gupta:PLB597'2004,Aarts:JHEP1502'2015,Ding:PRD94'2016} etc. 
However, we have used the kinetic theory approach by solving the 
relativistic Boltzmann transport equation in the relaxation-time 
approximation to calculate the electrical and thermal conductivities, 
where the interactions among particles are incorporated through their 
effective masses in the quasiparticle model at finite temperature, strong 
magnetic field and finite chemical potential. 

The rest of this paper is organized as follows. In section 2, 
we have first revisited the charge transport properties by 
calculating the electrical conductivity for an isotropic thermal 
medium at finite chemical potential and then calculated the same in an 
ambience of strong magnetic field. Similarly in section 3, after 
revisiting the thermal transport properties for an isotropic thermal 
medium at finite chemical potential by determining the thermal conductivity, 
we have calculated it in the presence of strong magnetic field. In the 
similar environment, we have studied some applications of the charge 
and thermal transport properties, {\em viz.}, the Knudsen number 
and the Wiedemann-Franz law in section 4. In section 5, we have 
determined the quasiparticle mass of quark in the presence of strong 
magnetic field and finite chemical potential. In section 6, we have 
discussed our results regarding electrical conductivity, thermal 
conductivity, Knudsen number and Wiedemann-Franz law. Finally, in 
section 7, we have written the conclusions. 

\section{Charge transport properties}
In this section, we are going to study the charge transport 
properties. Subsections 2.1 and 2.2 are devoted to the 
calculations of electrical conductivity for an isotropic 
dense QCD medium in the absence of magnetic field and for 
a dense QCD medium in the presence of strong magnetic 
field, respectively. 

\subsection{Isotropic dense QCD medium in the absence of magnetic field}
When an external electric field disturbs the system infinitesimally, an 
electric current is induced, which is given by
\begin{eqnarray}\label{current}
J_\mu = \sum_f g_f \int\frac{d^3\rm{p}}{(2\pi)^3\omega_f}
p_\mu [q_f\delta f_f(x,p)+{\bar q_f}\delta \bar{f_f}(x,p)]
~,\end{eqnarray}
where `$f$' stands for flavor and here we have used $f=u,d,s$. 
In the above equation $g_f$, $q_f$ ($\bar q_f$) and $\delta f_f $ 
($\delta \bar{f_f}$) represent the degeneracy factor, electric 
charge and infinitesimal change in the distribution function for the 
quark (antiquark) of $f$th flavor, respectively. According to the 
Ohm's law, the spatial component of the four-current is the product 
of the electrical conductivity and the external electric field,
\begin{eqnarray}\label{Ohm's law}
\mathbf{J}=\ec\mathbf{E}
~.\end{eqnarray}
Thus by comparing equations \eqref{current} and \eqref{Ohm's law}, one can 
obtain the electrical conductivity. The infinitesimal disturbance 
$\delta f_f$ can be determined from the relativistic Boltzmann transport 
equation (RBTE), which is written in the relaxation time approximation (RTA) \cite{Crecignani:2002} as
\be\label{R.B.T.E.(1)}
p^\mu\frac{\partial f_f(x,p)}{\partial x^\mu}+q_f F^{\rho\sigma} 
p_\sigma \frac{\partial f_f(x,p)}{\partial p^\rho}=-\frac{p_\nu u^\nu}{\tau_f}\delta f_f(x,p)
~,\ee
where $f_f=\delta f_f+f_f^{\rm iso}$, $F^{\rho\sigma}$ is the 
electromagnetic field strength tensor and the relaxation time 
for quarks (antiquarks), $\tau_f$ ($\tau_{\bar{f}}$) in a thermal 
medium is given \cite{Hosoya:NPB250'1985} by
\begin{eqnarray}
\tau_{f(\bar{f})}=\frac{1}{5.1T\alpha_s^2\log\left(1/\alpha_s\right)\left[1+0.12(2N_f+1)\right]}
~.\end{eqnarray}
The isotropic distribution functions for quark and antiquark of $f$th 
flavor are written as
\be\label{D.F.}
&&f_f^{\rm iso}=\frac{1}{e^{\beta(\omega_f-\mu_f)}+1}~, \\ 
&&\label{D.F.1}\bar{f_f}^{\rm iso}=\frac{1}{e^{\beta(\omega_f+\mu_f)}+1}
~,\ee
respectively, where $\omega_f=\sqrt{\mathbf{p}^2+m_f^2}$, $T=\beta^{-1}$ and 
$\mu_f$ is the chemical potential of $f$th flavor. For response of electric 
field, we set $\rho=i$ and $\sigma=0$ and {\em vice versa} in the 
calculation, so, $F^{i0}=\mathbf{E}$ and $F^{0i}=-\mathbf{E}$. Thus, 
RBTE \eqref{R.B.T.E.(1)} turns out to be
\be
q_f\mathbf{E}\cdot\mathbf{p}\frac{\partial f_f^{\rm iso}}{\partial p_0}
+q_f p_0\mathbf{E}\cdot\frac{\partial f_f^{\rm iso}}{\partial \mathbf{p}}
=-\frac{p_0}{\tau_f}\delta f_f
~.\ee
Now solving we get $\delta f_f$ as
\be
\delta f_f=2q_f\tau_f\beta\frac{\mathbf{E}\cdot\mathbf{p}}{\omega_f}
f_f^{\rm iso}(1-f_f^{\rm iso})
~.\ee
Similarly, $\delta \bar{f_f}$ is obtained as
\be
\delta \bar{f_f}=2{\bar q_f}\tau_{\bar{f}}\beta\frac{\mathbf{E}\cdot\mathbf{p}}{\omega_f}
\bar{f_f}^{\rm iso}(1-\bar{f_f}^{\rm iso})
~.\ee
After substituting the values of $\delta f_f$ and $\delta \bar{f_f}$ in 
eq. (\ref{current}), we get the electrical conductivity for an 
isotropic dense QCD medium in the absence of magnetic field as
\be\label{I.E.C.}
\sigma_{\rm el}^{\rm iso}=\frac{\beta}{3\pi^2}\sum_f g_f q_f^2\int d{\rm p}~\frac{{\rm p}^4}{\omega_f^2} ~ \left[\tau_f f_f^{\rm iso}(1-f_f^{\rm iso})+\tau_{\bar{f}}\bar{f_f}^{\rm iso}(1-\bar{f_f}^{\rm iso})\right]
~.\ee

\subsection{Dense QCD medium in the presence of strong magnetic field}
When the thermal medium comes under the influence of a strong 
magnetic field, the quark momentum gets split into the 
components which are transverse and longitudinal to the 
direction of magnetic field (say, z or 3-direction). As a 
result, the dispersion relation for the quark of $f$th flavor takes the 
following form,
\begin{eqnarray}\label{dispersion relation}
\omega_{f,n}(p_L)=\sqrt{p_L^2+2n\left|q_fB\right|+m_f^2}
~,\end{eqnarray}
where $n=0$, $1$, $2$,$\cdots$ represent the Landau levels. In the 
SMF limit ($|q_fB| \gg T^2$), the quarks only occupy the lowest Landau 
level ($n=0$), because they could not be excited thermally to the higher 
Landau levels due to very large energy gap 
$\sim\mathcal{O}(\sqrt{|q_fB|})$. Therefore $p_T\ll p_L$, which results 
in a momentum anisotropy and the distribution functions for 
quark and antiquark are written as
\be\label{A.D.F.(eB)}
&&f^B_f=\frac{1}{e^{\beta(\omega_f-\mu_f)}+1} ~, \\
&&\label{A.D.F.(eB1)}\bar{f}^B_f=\frac{1}{e^{\beta(\omega_f+\mu_f)}+1}
~,\ee
respectively. Here $\omega_f=\sqrt{p_3^2+m_f^2}$ according to the dispersion 
relation in the strong magnetic field limit, where the quark 
momentum becomes purely longitudinal \cite{Gusynin:PLB450'1999}. 
So, in the strong magnetic field regime, when an external 
electric field disturbs the system infinitesimally, an 
electromagnetic current is induced along the longitudinal 
direction (3-direction), which is given by
\begin{eqnarray}\label{current(eB1)}
J_3 = \sum_f g_f \int\frac{d^3\rm{p}}{(2\pi)^3\omega_f}
p_3 [q_f\delta f_f(\tilde{x},\tilde{p})+{\bar q_f}\delta \bar{f_f}(\tilde{x},\tilde{p})]
~,\end{eqnarray}
where $\tilde{x}=(x_0,0,0,x_3)$ and $\tilde{p}=(p_0,0,0,p_3)$. The 
(integration) phase factor also gets modified in terms of magnetic 
field \cite{Gusynin:NPB462'1996,Bruckmann:PRD96'2017} as $\int\frac{d^3{\rm p}}{(2\pi)^3}\rightarrow\frac{|q_fB|}{2\pi}\int \frac{dp_3}{2\pi}$. 
In the strong magnetic field limit, the electrical conductivity can 
be determined from the third component of current in Ohm's law,
\be\label{current(eB)}
J_3=\sigma_{\rm el} E_3
~.\ee
The infinitesimal disturbance ($\delta f_f$) can be evaluated from the 
relativistic Boltzmann transport equation in the relaxation time 
approximation, in conjunction with the strong magnetic field limit,
\begin{eqnarray}\label{R.B.T.E.(eB)}
p^0\frac{\partial f_f}{\partial x^0}+p^3\frac{\partial f_f}{\partial x^3}+q_f F^{\rho\sigma}p_\sigma \frac{\partial f_f}{\partial p^\rho}=-\frac{p_0}{\tau^B_f}\delta f_f
~,\end{eqnarray}
where the relaxation time, $\tau^B_{f(\bar{f})}$ in the strong magnetic field 
regime is given \cite{Hattori:PRD95'2017} by
\begin{eqnarray}
\tau^B_{f(\bar{f})}=\frac{\omega_f\left(e^{\beta\omega_f}-1\right)}{\alpha_sC_2m_f^2\left(e^{\beta\omega_f}+1\right)}\frac{1}{\int dp^\prime_3\frac{1}{\omega^\prime_f\left(e^{\beta\omega^\prime_f}+1\right)}}
~,\end{eqnarray}
where $C_2$ denotes the Casimir factor. For the response 
of electric field in the presence of strong magnetic 
field, we have set $\rho=3$ and $\sigma=0$ and 
{\em vice versa}, so, $F^{30}=E_3$ and $F^{03}=-E_3$. Solving 
eq. \eqref{R.B.T.E.(eB)}, we get $\delta f_f$ as
\be
\delta f_f &=& \frac{2\tau^B_f \beta q_fE_3p_3}{\omega_f}f^B_f\left(1-f^B_f\right)
.\ee
Similarly for antiquark, $\delta \bar{f_f}$ is evaluated as
\be
\delta \bar{f_f} &=& \frac{2\tau^B_{\bar{f}} \beta \bar{q_f}E_3p_3}{\omega_f}\bar{f}^B_f\left(1-\bar{f}^B_f\right)
.\ee
Finally substituting $\delta f_f$ and $\delta \bar{f_f}$ in 
eq. \eqref{current(eB1)} and then 
comparing with eq. \eqref{current(eB)}, we have calculated the 
electrical conductivity in the presence of strong magnetic 
field at finite chemical potential as
\begin{eqnarray}\label{A.E.C.(eb)}
\sigma_{\rm el}^B &=& \frac{\beta}{2\pi^2}\sum_f g_f q_f^2~|q_fB|\int dp_3~\frac{p_3^2}{\omega_f^2} ~ \left[\tau_f^B f^B_f\left(1-f^B_f\right)+\tau^B_{\bar{f}}\bar{f}^B_f\left(1-\bar{f}^B_f\right)\right]
.\end{eqnarray}

\section{Thermal transport properties}
In this section, we are going to study the thermal transport 
properties. In subsections 3.1 and 3.2, we have calculated the thermal 
conductivity for an isotropic dense QCD medium in the absence of 
magnetic field and for a dense QCD medium in the presence of strong 
magnetic field, respectively. 

\subsection{Isotropic dense QCD medium in the absence of magnetic field}
In a system, the flow of heat is directly proportional to the temperature 
gradient and the proportionality factor is known as the thermal 
conductivity. The flow of heat is not continuous, rather it diffuses, 
depending on the thermal properties of the medium. Thus, through the study 
of thermal conductivity in a medium, one can get the information on how the 
heat flows in that medium and how it affects the hydrodynamic equilibrium 
of the system with finite chemical potential.

The difference between the energy diffusion and the enthalpy diffusion 
gives the heat flow four-vector as
\be\label{heat flow}
Q_\mu=\Delta_{\mu\alpha}T^{\alpha\beta}u_\beta-h\Delta_{\mu\alpha}N^\alpha
,\ee
where $\Delta_{\mu\alpha}$ is the projection operator, 
$\Delta_{\mu\alpha}=g_{\mu\alpha}-u_\mu u_\alpha$, $T^{\alpha\beta}$ 
is the energy-momentum tensor, $N^\alpha$ is the particle 
flow four-vector, $h$ is the enthalpy per particle, 
$h=(\varepsilon+P)/n$ with $\varepsilon$, $P$ and $n$ represent the 
energy density, the pressure and the particle number density, 
respectively. $N^\alpha$ and $T^{\alpha\beta}$ are also defined in terms 
of the distribution functions as
\be
&&N^\alpha=\sum_f g_f\int \frac{d^3{\rm p}}{(2\pi)^3\omega_f}p^\alpha \left[f_f(x,p)+\bar{f}_f(x,p)\right] ~ \label{P.F.F.}, \\ &&T^{\alpha\beta}=\sum_f g_f\int \frac{d^3{\rm p}}{(2\pi)^3\omega_f}p^\alpha p^\beta \left[f_f(x,p)+\bar{f}_f(x,p)\right] ~ \label{E.M.T.}
,\ee
respectively. From equations \eqref{P.F.F.} and 
\eqref{E.M.T.}, the particle number density, the 
energy density and the pressure can be obtained 
as $n=N^\alpha u_\alpha$, $\varepsilon=u_\alpha T^{\alpha\beta} u_\beta$ and $P=-\Delta_{\alpha\beta}T^{\alpha\beta}/3$, respectively. As heat flow 
four-vector in the rest frame of the heat bath is orthogonal to the 
fluid four-velocity, {\em i.e.} $Q_\mu u^\mu=0$, heat flow is spatial, 
which under the action of external disturbance can be written in terms 
of the infinitesimal changes in the distribution functions as
\be\label{heat1}
\mathbf{Q}=\sum_f g_f\int \frac{d^3{\rm p}}{(2\pi)^3} ~ \frac{\mathbf{p}}{\omega_f}\left[(\omega_f-h_f)\delta f_f(x,p)+(\omega_f-\bar{h}_f)\delta \bar{f}_f(x,p)\right]
~.\ee
The Navier-Stokes equation relates the heat flow with the thermal 
potential ($U=\mu/T$) \cite{Greif:PRE87'2013} as
\be\label{heat}
\nonumber Q_\mu &=& -\kappa\frac{nT^2}{\varepsilon+P}\nabla_\mu U \\ &=& \kappa\left[\nabla_\mu T - \frac{T}{\varepsilon+P}\nabla_\mu P\right] 
,\ee
where $\kappa$ is the thermal conductivity and $\nabla_\mu$ is the 
four-gradient, $\nabla_\mu=\partial_\mu-u_\mu u_\nu\partial^\nu$. 
In the local rest frame, the spatial component of the heat flow is 
written as
\be\label{heat2}
\mathbf{Q}=-\kappa\left[\frac{\partial T}{\partial\mathbf{x}}-\frac{T}{nh}\frac{\partial P}{\partial\mathbf{x}}\right]
.\ee
One can thus obtain the thermal conductivity ($\kappa$) by comparing 
equations (\ref{heat1}) and (\ref{heat2}). Expanding the 
distribution function in terms of the gradients of flow velocity 
and temperature, the relativistic Boltzmann transport equation 
(\ref{R.B.T.E.(1)}) can be written as
\be\label{eq1}
p^\mu\partial_\mu T\frac{\partial f_f}{\partial T}+p^\mu\partial_\mu(p^\nu u_\nu)\frac{\partial f_f}{\partial p^0}+q_f\left[F^{0j}p_j\frac{\partial f_f}{\partial p^0}+F^{j0}p_0\frac{\partial f_f}{\partial p^j}\right]=-\frac{p^\nu u_\nu}{\tau_f}\delta f_f
~,\ee
where $p_0=\omega_f-\mu_f$ and for very small $\mu_f$, it can be approximated 
as $p_0\approx\omega_f$. Using the following partial derivatives,
\be
&&\frac{\partial f_f^{\rm iso}}{\partial T}=\frac{p_0}{T^2}f_f^{\rm iso}(1-f_f^{\rm iso}), \\ && \frac{\partial f_f^{\rm iso}}{\partial p^0}=-\frac{1}{T}f_f^{\rm iso}(1-f_f^{\rm iso}), \\ && \frac{\partial f_f^{\rm iso}}{\partial p^j}=-\frac{p^j}{Tp_0}f_f^{\rm iso}(1-f_f^{\rm iso})
,\ee
we solve eq. (\ref{eq1}) and get the infinitesimal disturbance,
\be\label{delta}
\nonumber\delta f_f &=& \nonumber-\frac{\tau_f f_f^{\rm iso}(1-f_f^{\rm iso})}{T}\left[\frac{p_0}{T}\partial_0 T+\frac{1}{T}p^j\partial_j T+T\partial_0\left(\frac{\mu_f}{T}\right)+\frac{T}{p_0}p^j\partial_j\left(\frac{\mu_f}{T}\right)\right. \\ && \hspace{3.3 cm}\left.-p^\nu\partial_0 u_\nu -\frac{p^jp^\nu}{p_0}\partial_j u_\nu - \frac{2q_f}{p_0}\mathbf{E}\cdot\mathbf{p}\right]
.\ee
Using $\partial_j\left(\frac{\mu_f}{T}\right)=-\frac{h_f}{T^2}\left(\partial_jT-\frac{T}{nh_f}\partial_jP\right)$ and $\partial_0 u_\nu=\nabla_\nu P/(nh_f)$ from the 
energy-momentum conservation, we obtain $\delta f_f$ as
\be
\nonumber\delta f_f &=& -\frac{\tau_f f_f^{\rm iso}(1-f_f^{\rm iso})}{T}\left[\frac{p_0}{T}\partial_0 T+\left(\frac{p_0-h_f}{p_0}\right)\frac{p^j}{T}\left(\partial_jT-\frac{T}{nh_f}\partial_jP\right)+T\partial_0\left(\frac{\mu_f}{T}\right)\right. \\ && \hspace{3.3 cm}\left.-\frac{p^jp^\nu}{p_0}\partial_j u_\nu - \frac{2q_f}{p_0}\mathbf{E}\cdot\mathbf{p}\right]
.\ee
Similarly $\delta \bar{f}_f$ is calculated as
\be
\nonumber\delta \bar{f}_f &=& -\frac{\tau_{\bar{f}} \bar{f}_f^{\rm iso}(1-\bar{f}_f^{\rm iso})}{T}\left[\frac{p_0}{T}\partial_0 T+\left(\frac{p_0-\bar{h}_f}{p_0}\right)\frac{p^j}{T}\left(\partial_jT-\frac{T}{n\bar{h}_f}\partial_jP\right)-T\partial_0\left(\frac{\mu_f}{T}\right)\right. \\ && \hspace{3.3 cm}\left.-\frac{p^jp^\nu}{p_0}\partial_j u_\nu - \frac{2\bar{q}_f}{p_0}\mathbf{E}\cdot\mathbf{p}\right]
.\ee
Then substituting the expressions of $\delta f_f$ and 
$\delta \bar{f}_f$ in eq. (\ref{heat1}) and comparing 
with eq. (\ref{heat2}), the thermal conductivity for the 
isotropic dense QCD medium in the absence of magnetic 
field is obtained as
\be\label{I.T.C.}
\kappa^{\rm iso}=\frac{\beta^2}{6\pi^2}\sum_fg_f\int d{\rm p} \frac{{\rm p}^4}{\omega_f^2} ~ \left[\tau_f(\omega_f-h_f)^2 ~ f_f^{\rm iso}(1-f_f^{\rm iso})+\tau_{\bar{f}}(\omega_f-\bar{h}_f)^2 ~ \bar{f}_f^{\rm iso}(1-\bar{f}_f^{\rm iso})\right]
.\ee

\subsection{Dense QCD medium in the presence of strong magnetic field}
The presence of strong magnetic field reduces the dynamics of 
quarks from three spatial dimensions to one spatial dimension, 
as a result, they can move only along the direction of magnetic 
field. Thus, in the strong magnetic field regime, the spatial 
component of heat flow is written as
\be\label{heat(eb1)}
Q_3=\sum_f\frac{g_f|q_fB|}{4\pi^2}\int dp_3 ~ \frac{p_3}{\omega_f}\left[(\omega_f-h_f^B) \delta f_f(\tilde{x},\tilde{p})+(\omega_f-\bar{h}_f^B) \delta \bar{f}_f(\tilde{x},\tilde{p})\right]
~.\ee
In addition, eq. (\ref{heat2}) also gets modified into
\be\label{heat(eb2)}
\nonumber Q_3 &=& -\kappa^B\left[\frac{\partial T}{\partial x_3}-\frac{T}{nh^B}\frac{\partial P}{\partial x_3}\right] \\ &=& \kappa^B\left[\partial_3T-\frac{T}{nh^B}\partial_3P\right]
.\ee
Here $h^B=(\varepsilon+P)/n$ denotes the enthalpy per 
particle in the presence of strong magnetic field. In this 
regime, the particle number density ($n$), the energy 
density ($\varepsilon$) and the pressure ($P$) are 
obtained from the following particle flow four-vector and 
energy-momentum tensor,
\begin{eqnarray}
N^\mu &=& \sum_f\frac{g_f|q_fB|}{4\pi^2}\int dp_3 \frac{\tilde{p}^\mu}{\omega_f} \left[f_f(\tilde{x},\tilde{p})+\bar{f}_f(\tilde{x},\tilde{p})\right], \label{P.F.F.(eb)} \\ 
T^{\mu\nu} &=& \sum_f\frac{g_f|q_fB|}{4\pi^2}\int dp_3 \frac{\tilde{p}^\mu\tilde{p}^\nu}{\omega_f} \left[f_f(\tilde{x},\tilde{p})+\bar{f}_f(\tilde{x},\tilde{p})\right] \label{E.M.T.(eb)} 
,\end{eqnarray}
respectively. The RBTE (\ref{R.B.T.E.(eB)}), in terms of the gradients 
of flow velocity and temperature is written as
\be\label{eq2}
\tilde{p}^\mu\frac{\partial T}{\partial \tilde{x}^\mu}\frac{\partial f_f}{\partial T}+\tilde{p}^\mu\frac{\partial (\tilde{p}^\nu u_\nu)}{\partial \tilde{x}^\mu}\frac{\partial f_f}{\partial p^0}+q_f\left[F^{03}p_3\frac{\partial f_f}{\partial p^0}+F^{30}p_0\frac{\partial f_f}{\partial p^3}\right]=-\frac{\tilde{p}^\nu u_\nu}{\tau_f^B}\delta f_f
~,\ee
where $\tilde{p}^\mu=(p^0,0,0,p^3)$ and $\tilde{x}^\mu=(x^0,0,0,x^3)$ are 
applicable for the calculation in strong magnetic field limit. Now 
using the following partial derivatives,
\be
&&\frac{\partial f^B_f}{\partial T}=\frac{p_0}{T^2}f^B_f\left(1-f^B_f\right), \\ && \frac{\partial f^B_f}{\partial p^0}=-\frac{1}{T}f^B_f\left(1-f^B_f\right), \\ && \frac{\partial f^B_f}{\partial p^3}=-\frac{p^3}{Tp_0}f^B_f\left(1-f^B_f\right)
,\ee
we get $\delta f_f$ from eq. (\ref{eq2}) as
\be
\nonumber\delta f_f &=& -\frac{\tau_f^B f^B_f(1-f^B_f)}{T}\left[\frac{p_0}{T}\partial_0 T+\left(\frac{p_0-h_f^B}{p_0}\right)\frac{p^3}{T}\left(\partial_3T-\frac{T}{nh_f^B}\partial_3P\right)+T\partial_0\left(\frac{\mu_f}{T}\right)\right. \\ && \left.-\frac{p^3\tilde{p}^\nu}{p_0}\partial_3 u_\nu-\frac{2q_f}{p_0}E_3p_3\right]
.\ee
Similarly $\delta \bar{f}_f$ is evaluated as
\be
\nonumber\delta \bar{f}_f &=& -\frac{\tau_{\bar{f}}^B \bar{f}^B_f(1-\bar{f}^B_f)}{T}\left[\frac{p_0}{T}\partial_0 T+\left(\frac{p_0-\bar{h}_f^B}{p_0}\right)\frac{p^3}{T}\left(\partial_3T-\frac{T}{n\bar{h}_f^B}\partial_3P\right)-T\partial_0\left(\frac{\mu_f}{T}\right)\right. \\ && \left.-\frac{p^3\tilde{p}^\nu}{p_0}\partial_3 u_\nu-\frac{2\bar{q}_f}{p_0}E_3p_3\right]
.\ee
Substituting $\delta f_f$ and $\delta \bar{f}_f$ in eq. (\ref{heat(eb1)}) 
and then comparing with eq. \eqref{heat(eb2)}, we get the thermal 
conductivity in the presence of strong magnetic field at finite chemical potential as
\begin{eqnarray}\label{A.T.C.(eb)}
\kappa^B=\frac{\beta^2}{4\pi^2}\sum_fg_f|q_fB|\int dp_3\frac{p_3^2}{\omega_f^2} ~ \left[\tau_f^B(\omega_f-h_f^B)^2f^B_f(1-f^B_f)+\tau_{\bar{f}}^B(\omega_f-\bar{h}_f^B)^2\bar{f}^B_f(1-\bar{f}^B_f)\right]
.\end{eqnarray}

\section{Applications}
This section is devoted to study some applications of the 
electrical and thermal conductivities. In subsection 4.1, 
we will observe the local equilibrium property of the 
medium through the Knudsen number in the presence of both strong 
magnetic field and finite chemical potential. In 
subsection 4.2, we will observe the relative behavior 
between the electrical conductivity and the thermal conductivity through the 
Wiedemann-Franz law for a thermal QCD medium in the aforesaid regime.

\subsection{Knudsen number}
The local equilibrium of a medium can be 
understood through the Knudsen number and 
it is defined by the ratio of mean free 
path ($\lambda$) to the characteristic 
length scale ($l$) of the medium,
\begin{eqnarray}
\Omega=\frac{\lambda}{l}
~.\end{eqnarray}
For the equilibrium hydrodynamics to be 
valid, the Knudsen number needs to be 
smaller than 1 or the mean free path 
requires to be smaller than the 
characteristic length scale of the medium. The 
mean free path is in turn related to the thermal 
conductivity ($\kappa$) by the following relation,
\begin{eqnarray}
\lambda=\frac{3\kappa}{vC_V}
~,\end{eqnarray}
where $C_V$ and $v$ are the specific heat at constant 
volume and the relative speed, respectively. So, $\Omega$ 
is now written as
\be
\Omega=\frac{3\kappa}{lvC_V}
~.\ee
In our calculation, we have set $v\simeq 1$, $l=4$ fm, and 
$C_V$ is determined from the energy-momentum tensor 
as $C_V=\partial (u_\mu T^{\mu\nu}u_\nu)/\partial T$. Our 
observation on the Knudsen number for a hot QCD matter in the 
presence of both strong magnetic field and finite chemical 
potential with the quasiparticle model has been described in subsection 6.3. 

\subsection{Wiedemann-Franz law}
The Wiedemann-Franz law helps us to understand the relation 
between the charge transport and the thermal transport 
in a system. This law states that the ratio of charged 
particle contribution of the thermal conductivity 
($\kappa$) to the electrical conductivity 
($\ec$) is equal to the product of Lorenz number ($L$) 
and temperature,
\begin{eqnarray}
\frac{\kappa}{\sigma_{\rm el}}=LT
~.\end{eqnarray}
For some metals which are good conductors of both electricity 
and heat, the ratio of thermal conductivity to 
electrical conductivity is proportional to temperature, 
{\em i.e.} $\frac{\kappa}{\sigma_{\rm el}} \propto T$, 
so the proportionality factor or the Lorenz number 
does not vary with the temperature, thus the Wiedemann-Franz 
law is perfectly satisfied. But some other systems report the 
violation of the Wiedemann-Franz law. In the thermally populated 
electron-hole plasma in graphene \cite{Crossno:S351'2016}, the 
violation is observed at high temperature, where the thermal conductivity 
is over an order of magnitude larger than the value expected for a 
Fermi liquid theory. In case of the two-flavor quark matter in the Nambu-Jona-Lasinio 
model \cite{Harutyunyan:PRD95'2017}, the violation 
occurs at temperatures $T \geq 250$ MeV due to the fact that the 
electrical conductivity remains nearly constant whereas the thermal 
conductivity shows decreasing behavior. For a strongly interacting 
QGP medium \cite{Mitra:PRD96'2017}, the Wiedemann-Franz 
law is violated at temperatures $T \leq 600$ MeV due to excessive rise 
of thermal conductivity as compared to electrical conductivity. The 
observations in unitary Fermi gas \cite{Husmann:PNAS115'2018,Han:PRA100'2019} 
indicate the violation of the Wiedemann-Franz law due to large charge 
conductance and small thermal conductance around the unitary regime 
containing large fraction of the preformed Cooper pairs. In case of the hot 
hadronic matter using the nonextensive 
statistics \cite{Rath:EPJA55'2019}, the Lorenz number has been 
found to increase with the temperature, thus confirming the violation 
of the Wiedemann-Franz Law. Our observation on the Wiedemann-Franz law 
for a hot QCD matter in the presence of both strong magnetic field 
and finite chemical potential with the quasiparticle model has 
been described in subsection 6.4. 

\section{Quasiparticle model in the presence of strong magnetic field and finite chemical potential}
In a thermal medium, the particles generally acquire thermally 
generated masses, known as quasiparticle masses. This mass arises 
due to the interaction of the particle with the other particles 
of the medium. Thus, in the quasiparticle model (QPM), QGP is treated 
as a system containing massive noninteracting quasiparticles. For 
a pure thermal medium, quasiparticle mass is temperature 
dependent, whereas for a thermal medium in the presence of strong 
magnetic field and chemical potential ($\mu_f$), quasiparticle mass 
becomes temperature, magnetic field and chemical potential 
dependent. The quasiparticle mass has been derived previously in 
different approaches, {\em viz.}, the Nambu-Jona-Lasinio (NJL) 
and polyakov NJL based quasiparticle models 
\cite{Fukushima:PLB591'2004,Ghosh:PRD73'2006,Abuki:PLB676'2009}, 
quasiparticle model with Gribov-Zwanziger quantization 
\cite{Su:PRL114'2015,Florkowski:PRC94'2016}, thermodynamically 
consistent quasiparticle model \cite{Bannur:JHEP0709'2007} etc. 
In a pure thermal medium, the effective mass (squared) of quark at 
finite $\mu_f$ is a well known result, which, for small 
chemical potential, is given up to $\mathcal{O}\left(\mu_f^2\right)$ 
\cite{Braaten:PRD45'1992,Peshier:PRD66'2002} by
\be\label{Q.P.M.}
m_{fT}^2=\frac{g^{\prime2}T^2}{6}\left(1+\frac{\mu_f^2}{\pi^2T^2}\right)
,\ee
where $g^\prime$ is the running coupling at finite temperature, 
finite chemical potential and zero magnetic field. In an ambience 
of strong magnetic field, the effective mass of quark can be 
determined from the effective quark propagator in the 
$p_0=0, p_z\rightarrow 0$ limit. The effective quark propagator is 
evaluated from the self-consistent Schwinger-Dyson equation in a strong 
magnetic field,
\be
S^{-1}(p_\parallel)=\gamma^\mu p_{\parallel\mu}-\Sigma(p_\parallel)
~.\ee
Thus, one requires to calculate the quark self-energy in the strong 
magnetic field regime, which is given by
\begin{eqnarray}\label{Q.S.E.}
\Sigma(p)=-\frac{4}{3} g^{2}i\int{\frac{d^4k}{(2\pi)^4}}\left[\gamma_\mu {S(k)}\gamma^\mu{D(p-k)}\right]
,\end{eqnarray}
where $g$ is the running coupling in the presence of a strong 
magnetic field \cite{Andreichikov:PRL110'2013,Ferrer:PRD91'2015,
Ayala:PRD98'2018,Viscosities}. The 
quark propagator $S(k)$ in vacuum in the strong magnetic field limit 
is given \cite{Schwinger:PR82'1951,Tsai:PRD10'1974} by the 
Schwinger proper-time method in momentum space, 
\be\label{q. propagator}
S(k)=ie^{-\frac{k^2_\perp}{|q_fB|}}\frac{\left(\gamma^0 k_0-\gamma^3 k_z+m_f\right)}{k^2_\parallel-m^2_f}\left(1-\gamma^0\gamma^3\gamma^5\right)
,\ee
where the following representations of the metric tensors and four vectors 
have been used,
\begin{eqnarray*}
&& g^{\mu\nu}_\perp={\rm{diag}}(0,-1,-1,0), ~~ g^{\mu\nu}_\parallel={\rm{diag}}(1,0,0,-1), \\ 
&& k_{\perp\mu}\equiv(0,k_x,k_y,0), ~~ k_{\parallel\mu}\equiv(k_0,0,0,k_z)
~.\end{eqnarray*}
Since gluon is an electrically neutral particle, its propagator in 
vacuum remains unaffected by the magnetic field and retains the 
form same as in the absence of magnetic field,
\be
\label{g. propagator}
D^{\mu \nu} (p-k)=\frac{ig^{\mu \nu}}{(p-k)^2}
~.\ee

Substituting the quark and gluon propagators in 
eq. \eqref{Q.S.E.}, the quark self-energy is 
determined using the imaginary-time formalism in the 
presence of strong magnetic field at finite 
chemical potential, where we have replaced the 
energy integral ($\int\frac{dp_0}{2\pi}$) by the 
sum over Matsubara frequencies and written the 
integration over the transverse component of the 
momentum in terms of $|q_fB|$. Thus, the quark 
self-energy \eqref{Q.S.E.} gets simplified into
\begin{eqnarray}\label{Q.S.E.(1)}
\nonumber\Sigma(p_\parallel) &=& \frac{2g^2}{3\pi^2}|q_fB|T\sum_n\int dk_z\frac{\left[\left(1+\gamma^0\gamma^3\gamma^5\right)\left(\gamma^0k_0
-\gamma^3k_z\right)-2m_f\right]}{\left[k_0^2-\omega^2_k\right]\left[(p_0-k_0)^2-\omega_{pk}^2\right]} \\ &=& \frac{2g^2|q_fB|}{3\pi^2}\int dk_z\left[(\gamma^0+\gamma^3\gamma^5)W^1-(\gamma^3+\gamma^0\gamma^5)k_zW^2\right]
,\end{eqnarray}
where $\omega^2_k=k_z^2+m_f^2$, $\omega_{pk}^2=(p_z-k_z)^2$ and 
$W^1$ and $W^2$ are the two frequency sums, which are given by
\be
&&W^1=T\sum_n \frac{k_0}{\left[k_0^2-\omega_k^2\right]\left[(p_0-k_0)^2-\omega_{pk}^2\right]} ~, \\ &&W^2=T\sum_n \frac{1}{\left[k_0^2-\omega_k^2\right]\left[(p_0-k_0)^2-\omega_{pk}^2\right]}
~.\ee
After calculating the above frequency sums and then substituting, 
we get the simplified form of the quark self-energy \eqref{Q.S.E.(1)} as
\begin{eqnarray}\label{Q.S.E.(2)}
\nonumber\Sigma(p_\parallel) &=& \frac{g^2|q_fB|}{3\pi^2}\int \frac{dk_z}{\omega_k}\left[\frac{1}{e^{\beta\omega_k}-1}+\frac{1}{2}\left\lbrace\frac{1}{e^{\beta(\omega_k+\mu_f)}+1}+\frac{1}{e^{\beta(\omega_k-\mu_f)}+1}\right\rbrace\right] \\ && \times\left[\frac{\gamma^0p_0+\gamma^3p_z}{p_\parallel^2}+\frac{\gamma^0\gamma^5p_z+\gamma^3\gamma^5p_0}{p_\parallel^2}\right]
,\end{eqnarray}
which after integration over $k_z$ takes the following 
approximated form for small chemical potential, 
\begin{eqnarray}\label{Q.S.E.(3)}
\nonumber\Sigma(p_\parallel) &\approx& \frac{g^2|q_fB|}{3\pi^2}\left[\frac{\pi T}{2m_f}-\ln(2)+\frac{7\mu_f^2\zeta(3)}{8\pi^2T^2}-\frac{31\mu_f^4\zeta(5)}{32\pi^4T^4}\right] \\ && \times\left[\frac{\gamma^0p_0}{p_\parallel^2}+\frac{\gamma^3p_z}{p_\parallel^2}+\frac{\gamma^0\gamma^5p_z}{p_\parallel^2}+\frac{\gamma^3\gamma^5p_0}{p_\parallel^2}\right]
,\end{eqnarray}
where $\zeta(s)$ is the Riemann zeta function with $s=3,5$ here. 
We also note that, we are working in the regime where $T>\mu_f$, so terms 
up to $\mathcal{O}\left(\frac{\mu_f^4}{T^4}\right)$ are kept in the above equation. 

The general covariant structure of the quark self-energy at 
finite temperature and magnetic field can be written \cite{Karmakar:PRD99'2019,Viscosities} as
\begin{equation}\label{general q.s.e.}
\Sigma(p_\parallel)=A\gamma^\mu u_\mu+B\gamma^\mu b_\mu+C\gamma^5\gamma^\mu u_\mu+D\gamma^5\gamma^\mu b_\mu
~,\end{equation}
where $A$, $B$, $C$ and $D$ represent the form factors, 
$u^\mu$ (1,0,0,0) denotes the preferred direction of 
heat bath which breaks the Lorentz symmetry and $b^\mu$ 
(0,0,0,-1) denotes the preferred direction of magnetic 
field which breaks the rotational symmetry. The form 
factors are evaluated as
\begin{eqnarray}
&&A=\frac{1}{4}{\rm Tr}\left[\Sigma\gamma^\mu u_\mu\right]=\frac{g^2|q_fB|}{3\pi^2}\left[\frac{\pi T}{2m_f}-\ln(2)+\frac{7\mu_f^2\zeta(3)}{8\pi^2T^2}-\frac{31\mu_f^4\zeta(5)}{32\pi^4T^4}\right]\frac{p_0}{p_\parallel^2} ~, \\ 
&&B=-\frac{1}{4}{\rm Tr}\left[\Sigma\gamma^\mu b_\mu\right]=\frac{g^2|q_fB|}{3\pi^2}\left[\frac{\pi T}{2m_f}-\ln(2)+\frac{7\mu_f^2\zeta(3)}{8\pi^2T^2}-\frac{31\mu_f^4\zeta(5)}{32\pi^4T^4}\right]\frac{p_z}{p_\parallel^2} ~, \\ 
&&C=\frac{1}{4}{\rm Tr}\left[\gamma^5\Sigma\gamma^\mu u_\mu\right]=-\frac{g^2|q_fB|}{3\pi^2}\left[\frac{\pi T}{2m_f}-\ln(2)+\frac{7\mu_f^2\zeta(3)}{8\pi^2T^2}-\frac{31\mu_f^4\zeta(5)}{32\pi^4T^4}\right]\frac{p_z}{p_\parallel^2} ~, \\ 
&&D=-\frac{1}{4}{\rm Tr}\left[\gamma^5\Sigma\gamma^\mu b_\mu\right]=-\frac{g^2|q_fB|}{3\pi^2}\left[\frac{\pi T}{2m_f}-\ln(2)+\frac{7\mu_f^2\zeta(3)}{8\pi^2T^2}-\frac{31\mu_f^4\zeta(5)}{32\pi^4T^4}\right]\frac{p_0}{p_\parallel^2}
~,\end{eqnarray}
where $C=-B$ and $D=-A$. In terms of the right-handed ($P_R=(1+\gamma^5)/2$) 
and left-handed ($P_L=(1-\gamma^5)/2$) chiral projection operators, the 
quark self-energy \eqref{general q.s.e.} is written as
\begin{equation}\label{projection}
\Sigma(p_\parallel)=P_R\left[(A+C)\gamma^\mu u_\mu+(B+D)\gamma^\mu b_\mu
\right]P_L+P_L\left[(A-C)\gamma^\mu u_\mu+(B-D)\gamma^\mu b_\mu\right]P_R
~,\end{equation}
which after the substitutions $C=-B$ and $D=-A$ gets simplified into
\begin{equation}\label{projection1}
\Sigma(p_\parallel)=P_R\left[(A-B)\gamma^\mu u_\mu+(B-A)\gamma^\mu b_\mu
\right]P_L+P_L\left[(A+B)\gamma^\mu u_\mu+(B+A)\gamma^\mu b_\mu\right]P_R
~.\end{equation}
Now with the quark self-energy \eqref{projection1}, the self-consistent 
Schwinger-Dyson equation in the presence of a strong magnetic field is 
written as
\be
\nonumber S^{-1}(p_\parallel) &=& \gamma^\mu p_{\parallel\mu}-\Sigma(p_\parallel) \\ &=& P_R\gamma^\mu X_\mu P_L+P_L\gamma^\mu Y_\mu P_R
~,\ee
where
\begin{eqnarray}
&&\gamma^\mu X_\mu=\gamma^\mu p_{\parallel\mu}-(A-B)\gamma^\mu u_\mu-(B-A)\gamma^\mu b_\mu ~, \\ 
&&\gamma^\mu Y_\mu=\gamma^\mu p_{\parallel\mu}-(A+B)\gamma^\mu u_\mu-(B+A)\gamma^\mu b_\mu
~.\end{eqnarray}
Thus, we get the effective quark propagator as
\be
S(p_\parallel)=\frac{1}{2}\left[P_R\frac{\gamma^\mu Y_\mu}{Y^2/2}P_L+
P_L\frac{\gamma^\mu X_\mu}{X^2/2}P_R\right]
,\ee
where
\begin{eqnarray}
&&\frac{X^2}{2}=X_1^2=\frac{1}{2}\left[p_0-(A-B)\right]^2-\frac{1}{2}\left[p_z+(B-A)\right]^2 ~, \\ 
&&\frac{Y^2}{2}=Y_1^2=\frac{1}{2}\left[p_0-(A+B)\right]^2-\frac{1}{2}\left[p_z+(B+A)\right]^2
~.\end{eqnarray}
The thermal mass (squared) at finite temperature and finite chemical 
potential in the presence of strong magnetic field is finally 
calculated by taking the $p_0=0, p_z\rightarrow 0$ limit of either 
$X_1^2$ or $Y_1^2$ (both of them are equal in this limit) as
\begin{eqnarray}\label{Mass}
m_{fT,B}^2=X_1^2\Big{|}_{p_0=0,p_z\rightarrow 0}=Y_1^2\Big{|}_{p_0=0,p_z\rightarrow 0}=\frac{g^2|q_fB|}{3\pi^2}\left[\frac{\pi T}{2m_f}-\ln(2)+\frac{7\mu_f^2\zeta(3)}{8\pi^2T^2}-\frac{31\mu_f^4\zeta(5)}{32\pi^4T^4}\right]
,\end{eqnarray}
which depends on temperature, chemical potential and magnetic field. 

In the calculation, we have chosen a specific range of temperature and 
magnetic field in such a way that the condition of strong magnetic 
field limit ($eB \gg T^2$) is satisfied. Thus, we have set the 
magnetic field at $15$ $m_{\pi}^2$ and the temperature in the 
range $0.16$ GeV - $0.4$ GeV. In addition, the chemical potential 
($\mu_f$) for all flavors are taken the same, {\em i.e.} $\mu_f=\mu$ 
and we have used the chemical potential, $\mu=0.06$ GeV, which is 
smaller than the temperature and the magnetic field. In the next 
section, we are going to discuss the results in quasiparticle model by 
using the temperature and chemical potential-dependent 
quasiparticle mass \eqref{Q.P.M.} for the isotropic dense thermal 
medium, and temperature, chemical potential and magnetic 
field-dependent quasiparticle mass \eqref{Mass} for the dense thermal 
medium in the presence of strong magnetic field. 

\begin{figure}[]
\begin{center}
\begin{tabular}{c c}
\includegraphics[width=6.3cm]{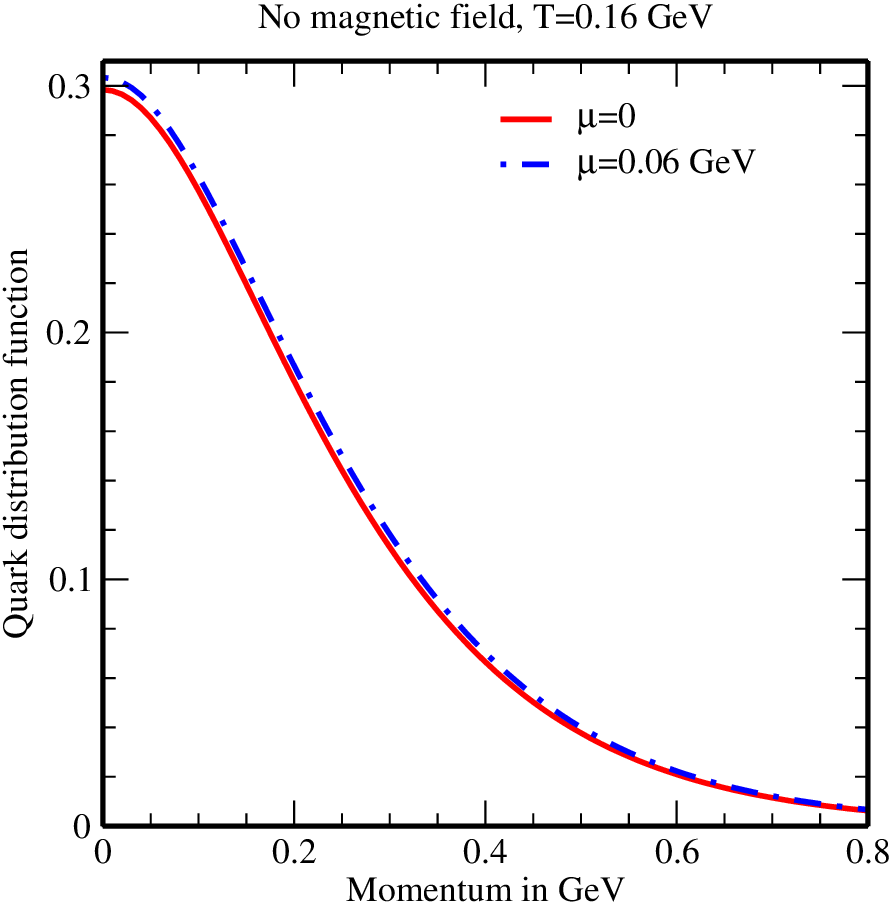}&
\hspace{0.423 cm}
\includegraphics[width=6.3cm]{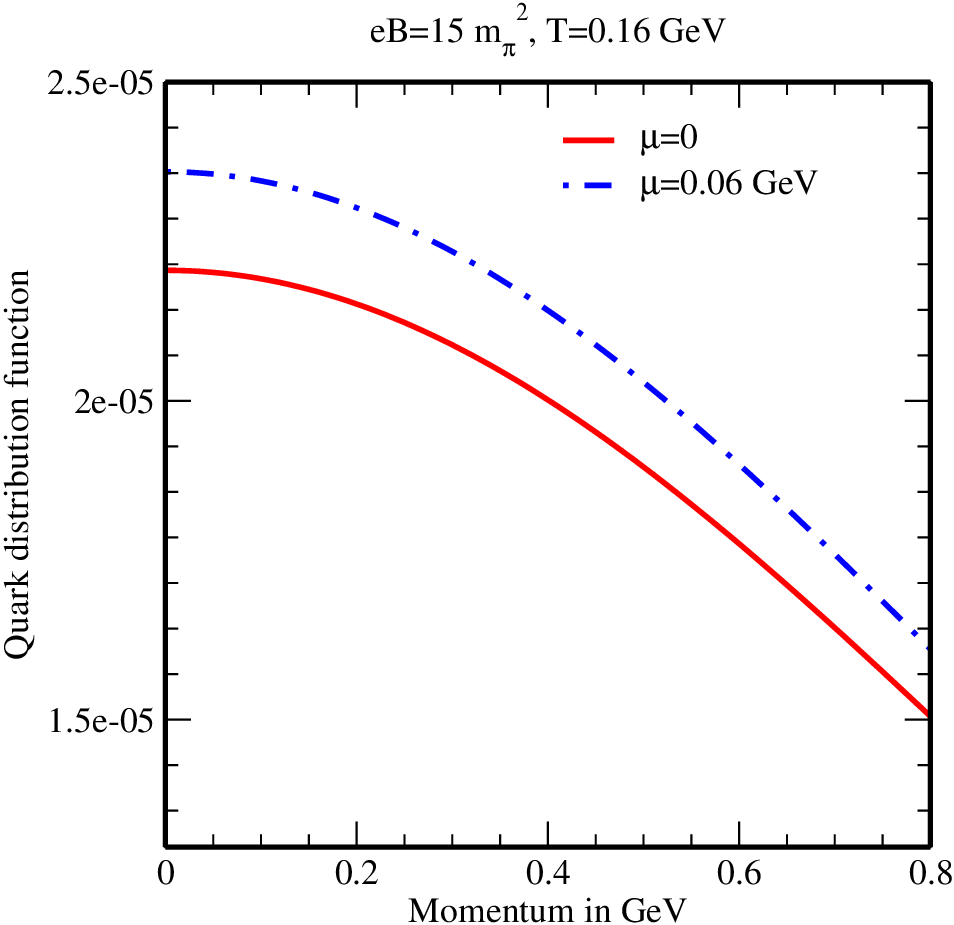} \\
a & b
\end{tabular}
\caption{Variations of the quark distribution function with 
momentum at low temperature in the (a) absence and (b) presence 
of strong magnetic field.}\label{fup.1}
\end{center}
\end{figure}

\begin{figure}[]
\begin{center}
\begin{tabular}{c c}
\includegraphics[width=6.3cm]{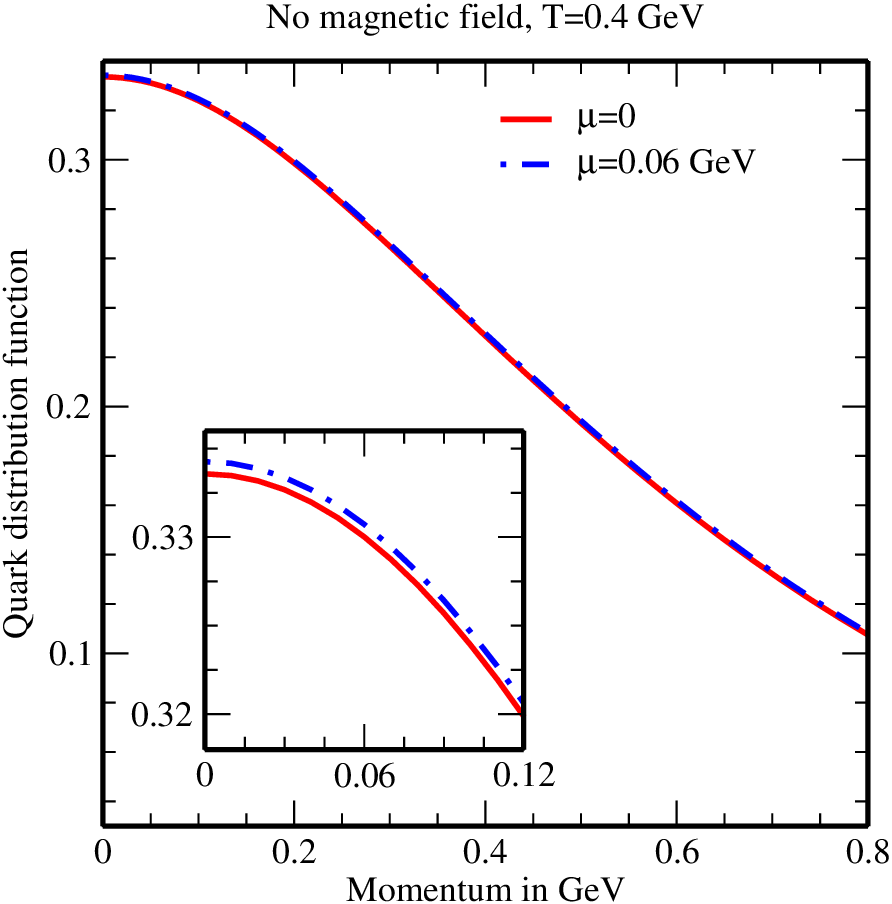}&
\hspace{0.423 cm}
\includegraphics[width=6.3cm]{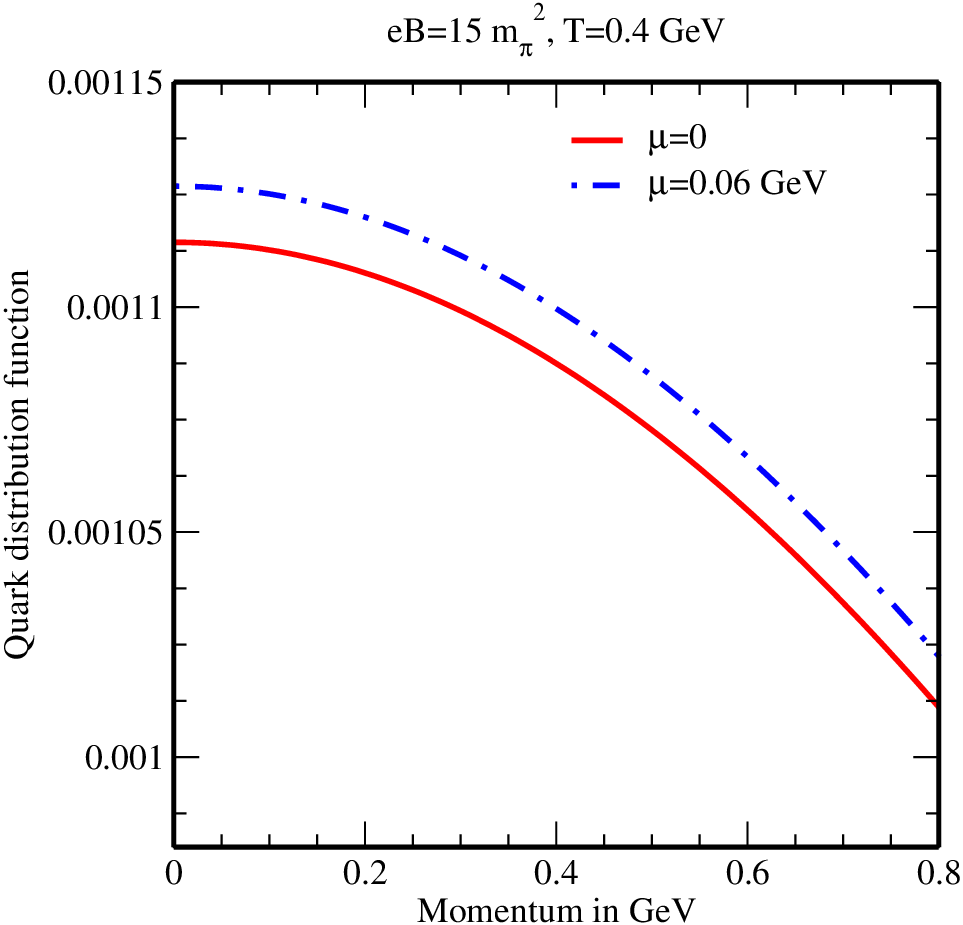} \\
a & b
\end{tabular}
\caption{Variations of the quark distribution function with 
momentum at high temperature in the (a) absence and (b) presence 
of strong magnetic field.}\label{fup.2}
\end{center}
\end{figure}

Before proceeding to discuss the results on the transport properties 
and their applications in the presence of strong magnetic field and 
finite chemical potential, it is necessary to discuss how the 
distribution function behaves in the similar environment, because in 
kinetic theory, the distribution function in general embraces most 
of the information about the properties of transport coefficients. In 
figure \ref{fup.1} and figure \ref{fup.2}, the distribution function 
for $u$ quark in the quasiparticle model has been plotted as a 
function of momentum at low temperature and high temperature, 
respectively. We have observed that, both in the absence and 
presence of strong magnetic field, the distribution function is 
larger at finite chemical potential in comparison to that at zero 
chemical potential. It is also clear that the distribution function 
has larger value at higher temperature (figure \ref{fup.2}) as 
compared to the low temperature case (figure \ref{fup.1}). 

\section{Results and discussions}
In this section, we are going to discuss the results regarding 
electrical conductivity, thermal conductivity, Knudsen number 
and Wiedemann-Franz law. 

\subsection{Electrical conductivity}
\begin{figure}[]
\begin{center}
\begin{tabular}{c c}
\includegraphics[width=6.3cm]{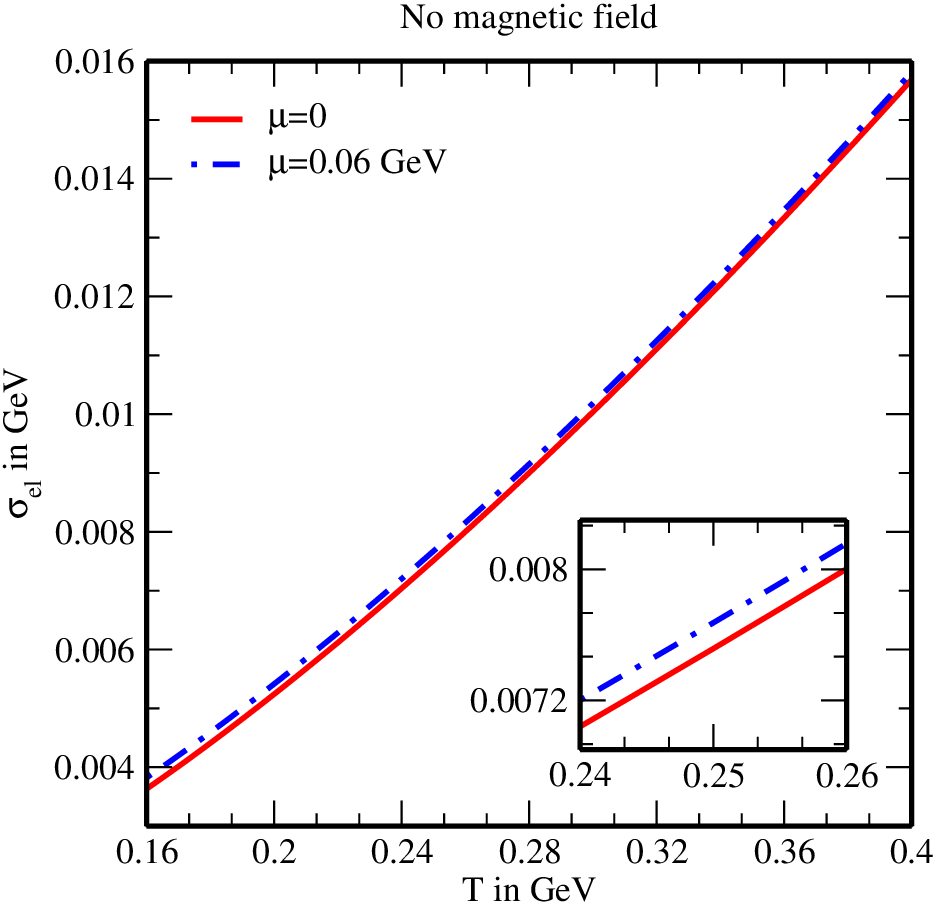}&
\hspace{0.74 cm}
\includegraphics[width=6.3cm]{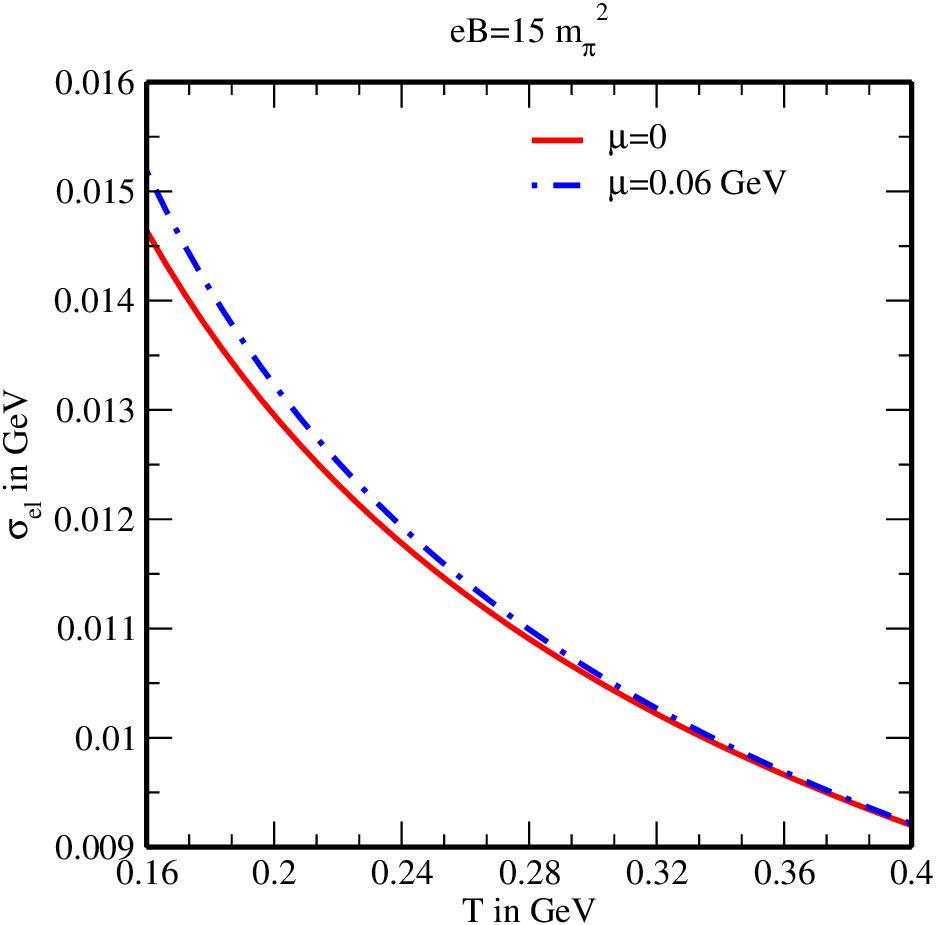} \\
a & b
\end{tabular}
\caption{Variations of the electrical conductivity with temperature 
in the (a) absence and (b) presence of strong magnetic field.}\label{Fig.1}
\end{center}
\end{figure}

From figure \ref{Fig.1}, we have observed that, the electrical conductivity 
gets enhanced in a strong magnetic field as compared to the zero 
magnetic field case. There are two main reasons behind this 
behavior of electrical conductivity, the first one is the 
distribution function and the second one is the dispersion 
relation. The strong magnetic field restricts the 
dynamics of charged particles to one spatial dimension 
(along the direction of magnetic field), 
which implies a stretching of the distribution 
function along the direction of magnetic field and a squeezing of the 
phase space, thus indicating a change in the dispersion relation. Therefore, the 
phase space integral is divided into longitudinal 
and transverse parts with respect to the direction of 
magnetic field, where the transverse part gives an 
overall factor of $|q_fB|$. In addition, the distribution 
functions also embody the effects of temperature and magnetic 
field through the effective masses in quasiparticle model. As a 
result, $\ec$ calculated in the strong 
magnetic field regime depends explicitly on magnetic field as 
well as temperature, whereas in the absence of magnetic field it 
depends only on temperature. According to the strong 
magnetic field (SMF) limit ($|q_fB| \gg T^2$), the energy scale 
associated with the magnetic field is much greater than the 
energy scale associated with the temperature, so $\ec$ becomes 
more sensitive to the magnetic field and thus electrical conductivity 
gets increased. 

But with the rise of temperature, electrical conductivity decreases 
(figure \ref{Fig.1}b), contrary to its increase in the absence of magnetic 
field (figure \ref{Fig.1}a). This behavior of $\ec$ with temperature can be 
understood from the fact that, in the strong magnetic field 
regime, temperature is the weak energy scale, so it may leave a 
reverse effect on $\ec$, whereas in a pure thermal medium 
({\em i.e.} in the absence of magnetic field), temperature is the dominant 
energy scale, so its effect on $\sigma_{\rm el}$ is more pronounced as 
compared to that in a strong magnetic field and the medium becomes 
electrically more conductive at higher temperatures. 

We have also found that the finite chemical potential enhances 
$\sigma_{\rm el}$ both in the absence of magnetic field 
(figure \ref{Fig.1}a) and in the presence of strong magnetic 
field (figure \ref{Fig.1}b). It is well known from the nonrelativistic 
Drude's formula that the electrical conductivity is directly 
proportional to the number density, which is the integration of 
distribution function over momentum space. In our case, the increase 
in the value of $\sigma_{\rm el}$ due to the finite chemical potential is also 
understood from the increase of distribution function at 
finite chemical potential (figures \ref{fup.1} and \ref{fup.2}). 

\subsection{Thermal conductivity}
\begin{figure}[]
\begin{center}
\begin{tabular}{c c}
\includegraphics[width=6.3cm]{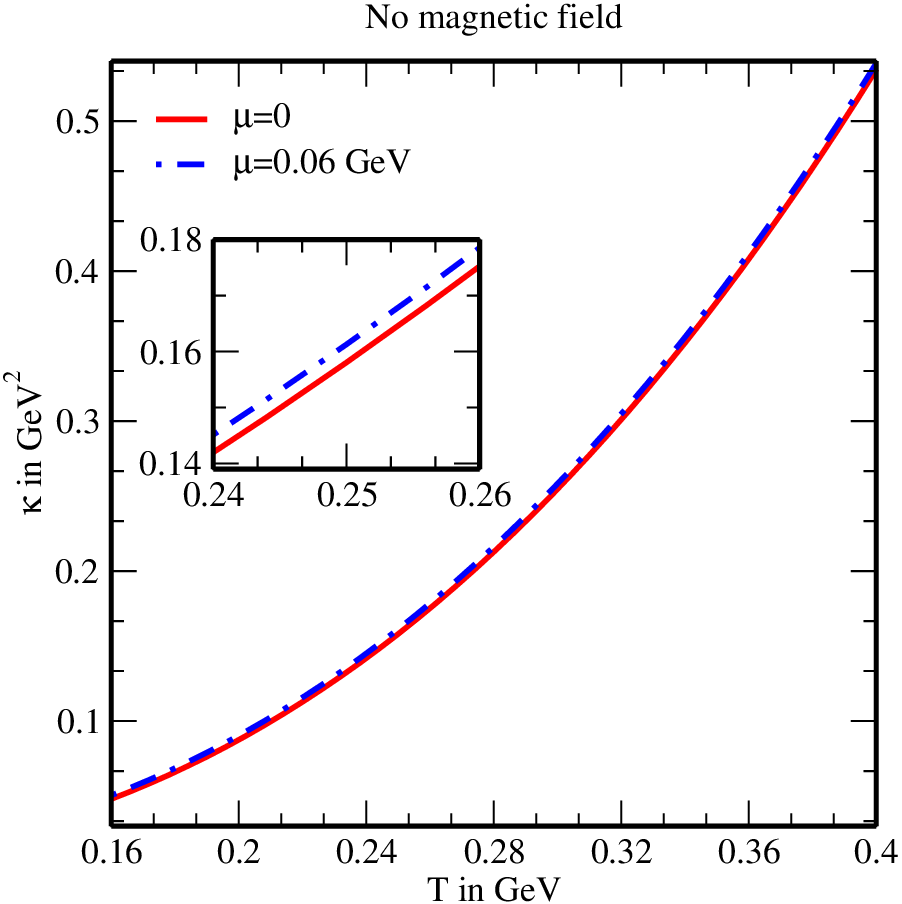}&
\hspace{0.74 cm}
\includegraphics[width=6.3cm]{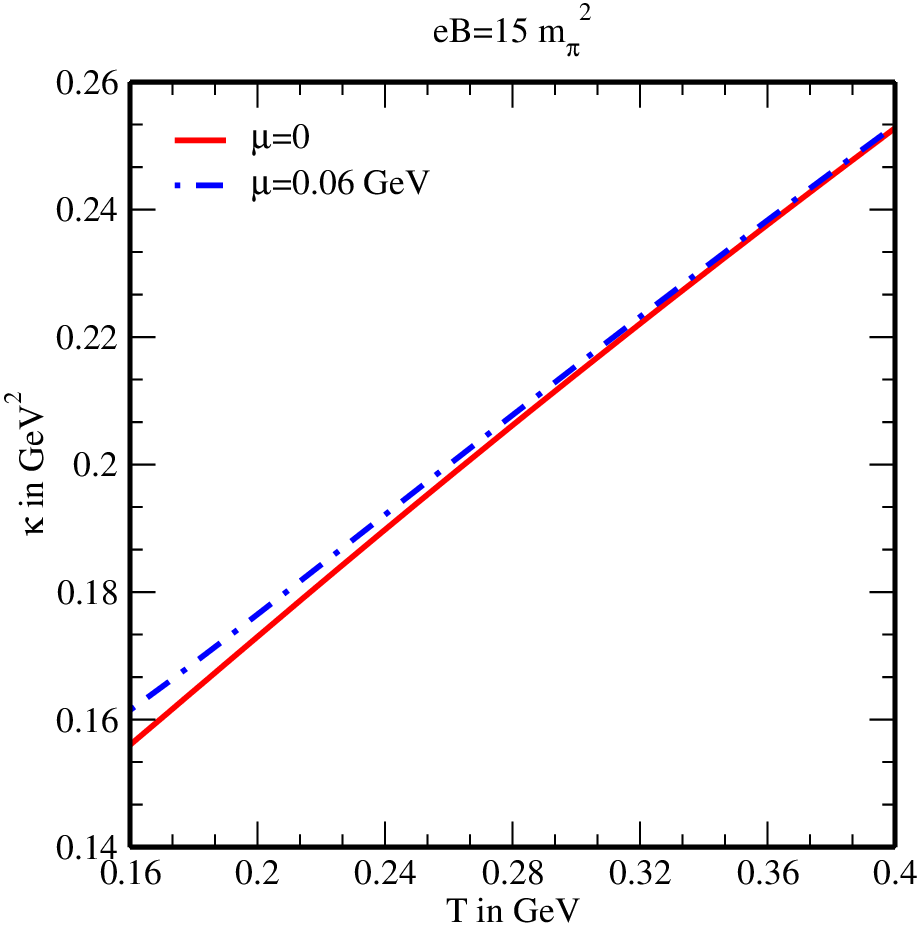} \\
a & b
\end{tabular}
\caption{Variations of the thermal conductivity with temperature 
in the (a) absence and (b) presence of strong magnetic field.}\label{Fig.2}
\end{center}
\end{figure}

Figure \ref{Fig.2} shows the variations of the thermal conductivity 
with the temperature in the absence and in the presence of strong 
magnetic field and chemical potential. The presence of strong 
magnetic field makes $\kappa$ increased as compared to the 
zero magnetic field case and it shows slow increasing trend 
with temperature (figure \ref{Fig.2}b), contrary to its rapid 
increasing trend in the isotropic medium in the absence of 
magnetic field (figure \ref{Fig.2}a). This behavior of 
$\kappa$ with temperature can be understood qualitatively 
from the fact that the temperature is the weak energy scale 
in the SMF limit ($|q_fB| \gg T^2$), so its effect on $\kappa$ 
is meager, whereas at zero magnetic field the temperature is the strong 
energy scale, so its effect on $\kappa$ is abundant. The existence 
of finite chemical potential rises the value of $\kappa$ and this rise 
is more pronounced in the presence of strong magnetic field. This 
difference in the behavior of thermal conductivity in the absence 
and presence of strong magnetic field and chemical potential is 
attributed to the differences in the behaviors of distribution 
functions and the dispersion relations in the absence and presence 
of magnetic field and chemical potential. 

\subsection{Knudsen number}
\begin{figure}[]
\begin{center}
\begin{tabular}{c c}
\includegraphics[width=6.3cm]{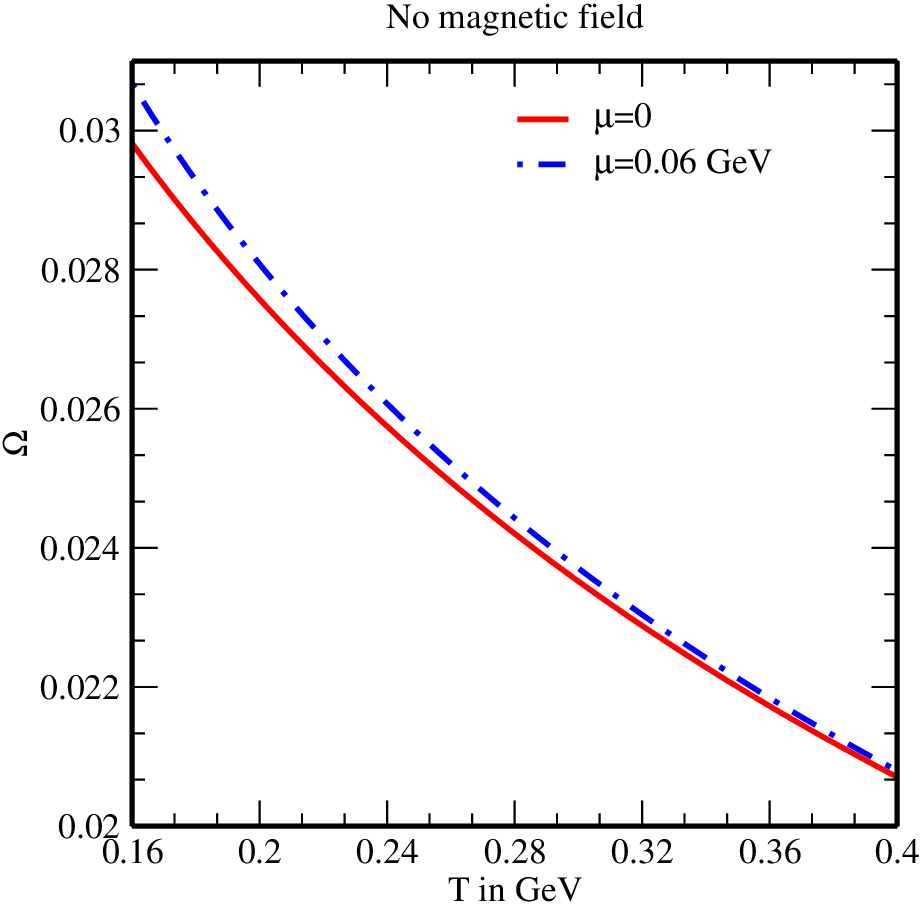}&
\hspace{0.74 cm}
\includegraphics[width=6.3cm]{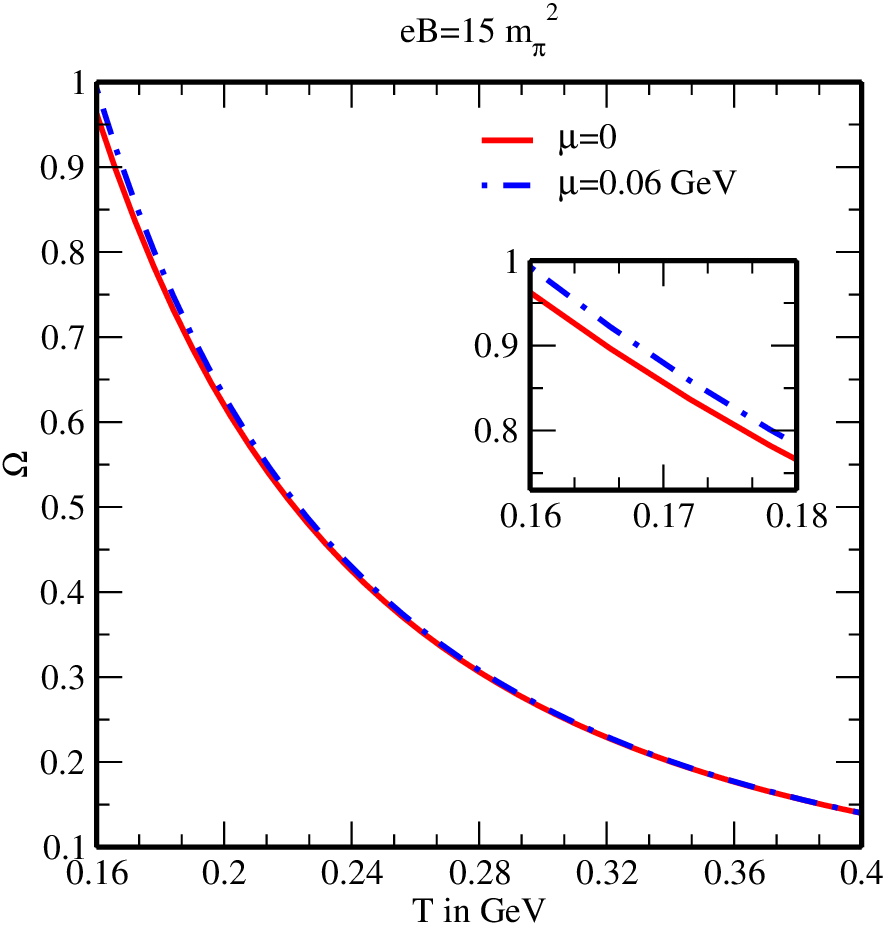} \\
a & b
\end{tabular}
\caption{Variations of the Knudsen number with temperature 
in the (a) absence and (b) presence of strong magnetic field.}\label{Fig.3}
\end{center}
\end{figure}

From figure \ref{Fig.3} we have found that, in the presence of strong 
magnetic field, the Knudsen number ($\Omega$) becomes larger than its 
value in the isotropic medium with no magnetic field. We have also noticed that, 
in the additional presence of chemical potential, the Knudsen number becomes 
further increased and approaches 1 near $T_c=0.16$ GeV (figure \ref{Fig.3}b), 
{\em i.e.} the mean free path gets closer to the size of the system, unlike the 
isotropic case in the absence of magnetic field where although $\Omega$ 
gets risen at finite chemical potential, but it remains much lower than unity 
(figure \ref{Fig.3}a). Thus, the medium moves slightly away from the 
local equilibrium at finite chemical potential in an ambience of strong 
magnetic field. 

\subsection{Wiedemann-Franz law}
\begin{figure}[]
\begin{center}
\begin{tabular}{c c}
\includegraphics[width=6.3cm]{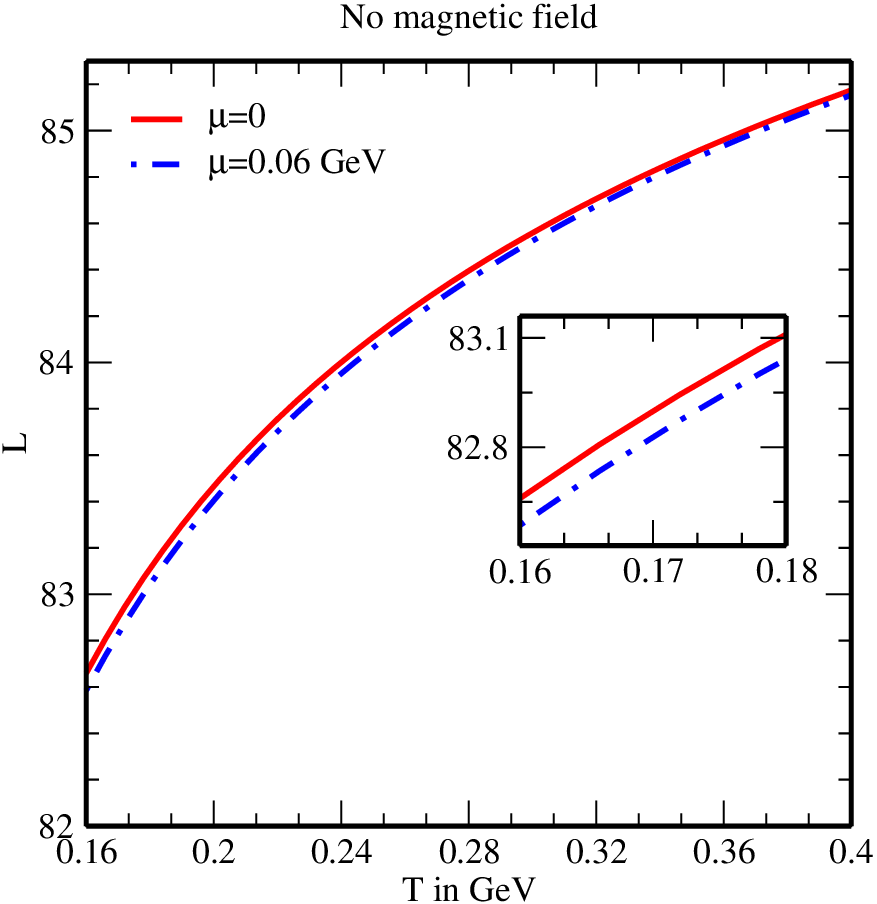}&
\hspace{0.74 cm}
\includegraphics[width=6.3cm]{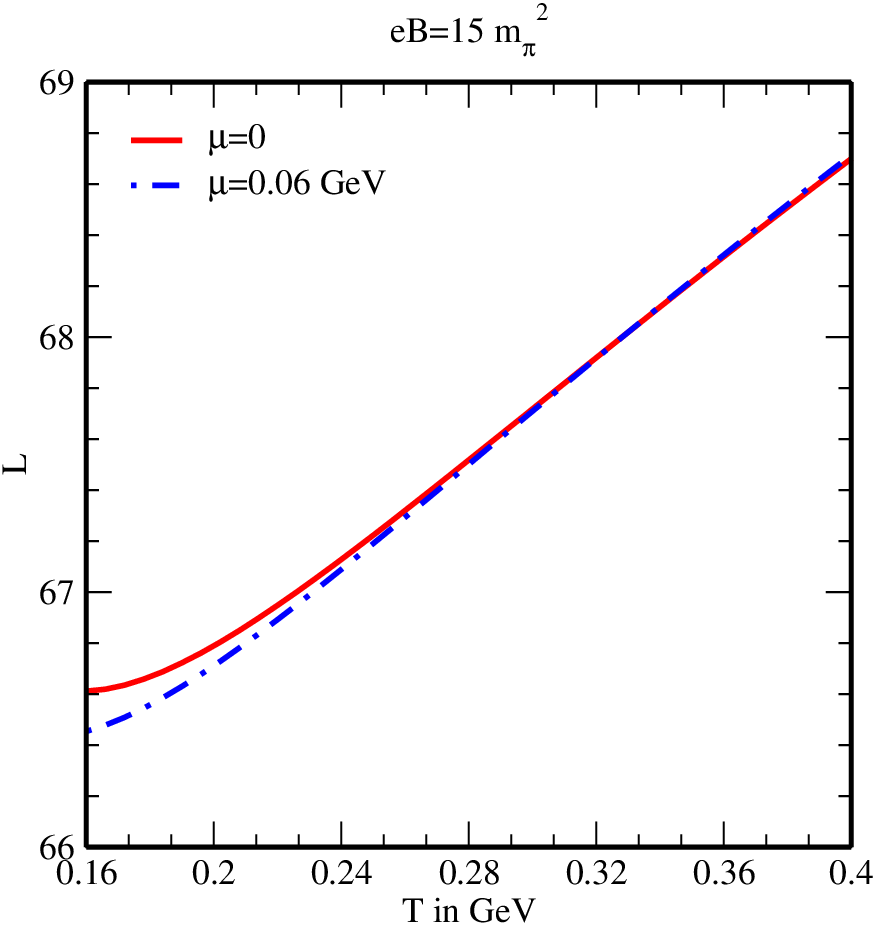} \\
a & b
\end{tabular}
\caption{Variations of the Lorenz number with temperature 
in the (a) absence and (b) presence of strong magnetic field.}\label{Fig.4}
\end{center}
\end{figure}

In figure \ref{Fig.4}, we have examined the validity of the Wiedemann-Franz law 
for a hot QCD matter in the presence of both strong magnetic field and 
finite chemical potential and also compared with the same in the 
absence of magnetic field. Figure \ref{Fig.4}a depicts the 
variation of the Lorenz number ($L$) with the temperature for isotropic 
medium in the absence of magnetic field and figure \ref{Fig.4}b depicts the same 
in the presence of strong magnetic field. We have noticed a decrease in the 
magnitude of $L$ due to the strong magnetic field as compared to the zero magnetic 
field case. In both the cases, finite chemical potential further reduces the 
magnitude of the Lorenz number, but it remains larger than unity. Thus for a hot 
QCD matter, thermal conductivity is much larger than its electrical conductivity 
at any fixed temperature; however, this difference between the thermal and 
electrical conducting characteristics becomes increased in the presence of a strong 
magnetic field and the additional presence of chemical potential further 
increases this difference, which also corroborate the observations 
on $\ec$ in figure \ref{Fig.1} and $\kappa$ in figure \ref{Fig.2}. 

We have also observed that, in all scenarios, {\em i.e.} in the absence 
as well as in the presence of strong magnetic field and chemical 
potential, with the increase of temperature, the Lorenz number does not 
remain constant, rather it increases. Thus, for a QCD matter, the 
dominance of thermal conductivity over electrical conductivity is more 
pronounced at higher temperatures and it explains that, the QCD matter 
is not a good conductor of electricity, but a very good conductor of 
heat, unlike some metals which are good conductors of both electricity 
and heat. In this perspective the QCD matter is very different from the 
metals. Therefore, in all abovementioned scenarios, the Wiedemann-Franz 
law gets violated for a hot QCD matter. 

\section{Conclusions}
In this work, we have studied how the strong magnetic 
field affects the charge and thermal transport properties 
of the hot QCD matter at finite chemical potential and 
observed the deviations from their isotropic counterparts 
in the isotropic medium in the absence of magnetic field. 
In calculating the electrical and thermal conductivities we 
have followed the kinetic theory approach in the relaxation time 
approximation, where the interactions are incorporated through the 
effective masses of particles at finite temperature, strong 
magnetic field and finite chemical potential in quasiparticle 
model. After revisiting the calculations of electrical and thermal 
conductivities for the isotropic dense medium, we determined these 
conductivities in the presence of strong magnetic field. We have 
observed that the values of electrical and thermal conductivities 
get increased in the presence of strong magnetic field in 
comparison to those in the isotropic medium at zero magnetic field, 
and the additional presence of chemical potential further 
increases their values. In applications of the aforesaid 
conductivities, we have investigated the local equilibrium 
property through the Knudsen number and the relative behavior 
between thermal conductivity and electrical conductivity 
through the Lorenz number in Wiedemann-Franz law in the presence 
of strong magnetic field and finite chemical potential. We have 
observed that the Knudsen number in the presence of strong 
magnetic field and finite chemical potential gets increased in 
comparison to the isotropic one at zero magnetic field and 
approaches 1 around $T_c=0.16$ GeV, but at higher temperatures, 
it becomes less than 1. Thus around $T_c$, the mean free path 
gets closer to the characteristic length scale of the system 
and the medium may move slightly away from the local 
equilibrium state, whereas at higher temperatures, the 
medium returns back to its equilibrium state. In the 
Wiedemann-Franz law, the Lorenz number ($\kappa/(\ec T)$) in the 
absence and also in the presence of strong magnetic field and 
chemical potential does not remain constant, rather it increases 
with the rise of temperature. So, this behavior of the Lorenz number 
indicates the violation of the Wiedemann-Franz law for a hot and 
dense QCD matter in the presence of strong magnetic field. 

Our observations on electrical and thermal conductivities are important 
from the phenomenological point of view: for large electrical 
conductivity, the lifetime of the strong magnetic field 
produced by the peripheral heavy ion collision is expected to be increased 
\cite{Tuchin(1):PRC88'2013,McLerran:NPA929'2014,Huang:RPP79'2016}. The 
emergence of chiral magnetic effect in the initial stage of the heavy ion 
collision is associated with the generation of current along the direction 
of magnetic field, which thus depends on the magnitude of the electrical 
conductivity \cite{Fukushima:PRD78'2008,Zhao:PPNP107'2019}. It was 
discussed that the dilepton 
production rate is directly proportional to the electrical conductivity 
of the quark-gluon plasma \cite{Moore:arXiv0607172}. Thus, the enhancement 
of electrical conductivity due to the strong magnetic field may imply 
large production rate of dilepton in experiments at ultrarelativistic 
heavy ion collisions. In addition, it was suggested that the 
anisotropy of the electrical conductivity helps in the production of the 
elliptic flow ($v_2$) of the photon \cite{Yin:PRC90'2014}. Our result also 
exhibits anisotropy due to the strong magnetic field, so it may 
significantly affect $v_2$ of the photon. Since the thermal conductivity is 
linked to the local equilibrium of the medium through the mean free path, 
its magnitude in a strong magnetic field can be used as an indicator to know 
how close the medium produced in peripheral heavy ion collision is to the 
equilibrium state in the presence of the strong magnetic field. The 
exploration of electrical and thermal conductivities of the hot and dense 
QCD matter in the strong magnetic field regime is also helpful to 
understand the conductive properties in other areas where strong magnetic 
fields might exist, such as the cores of the dense magnetars and the 
beginning of the universe, in addition to the ultrarelativistic heavy ion collisions. 

\section{Acknowledgment}
One of us (B. K. P.) is thankful to the Council of Scientific and 
Industrial Research (Grant No. 03(1407)/17/EMR-II) for the financial assistance.

\end{document}